\documentclass[english,aps,prl,notitlepage,twocolumn,superscriptaddress,longbibliography]{revtex4-1}
\usepackage[T1]{fontenc}
\usepackage[latin9]{inputenc}
\usepackage{verbatim}
\usepackage{amsmath}
\usepackage{graphicx}

\makeatletter
\usepackage[colorlinks,citecolor=blue,linkcolor=blue]{hyperref}

\makeatother

\usepackage{babel}
\begin{document}
\title{Achieve Higher Efficiency at Maximum Power with Finite-time Quantum
Otto Cycle}
\author{Jin-Fu Chen}
\address{Beijing Computational Science Research Center, Beijing 100193, China}
\address{Graduate School of China Academy of Engineering Physics, No. 10 Xibeiwang
East Road, Haidian District, Beijing, 100193, China}
\author{Chang-Pu Sun}
\address{Beijing Computational Science Research Center, Beijing 100193, China}
\address{Graduate School of China Academy of Engineering Physics, No. 10 Xibeiwang
East Road, Haidian District, Beijing, 100193, China}
\author{Hui Dong}
\email{hdong@gscaep.ac.cn}

\address{Graduate School of China Academy of Engineering Physics, No. 10 Xibeiwang
East Road, Haidian District, Beijing, 100193, China}
\date{\today}
\begin{abstract}
The optimization of heat engines was intensively explored to achieve
higher efficiency while maintaining the output power. However, most
investigations were limited to few finite-time cycles, e.g. Carnot-like
cycle, due to the complexity of the finite-time thermodynamics. In
this paper, we propose a new class of finite-time engine with quantum
Otto cycle, and demonstrate a higher achievable efficiency at maximum
power. The current model can be widely utilized benefited from the
general $\mathcal{C}/\tau^{2}$ scaling of extra work for finite-time
adiabatic process with long control time $\tau$. We apply the current
perturbation method to the quantum piston model and calculate the
efficiency at maximum power, which is validated with exact solution.
\end{abstract}
\maketitle

\section{Introduction}

The emergent studies of quantum thermodynamics \citep{Campisi_2011,Esposito_2009,Maruyama2009,Strasberg_2017,Vinjanampathy_2016}
have boosted the reminiscent investigation of heat engines into the
microscopic level, especially on the optimizing performance \citep{Curzon_1975,Geva1992,Esposito_2010,Ryabov2016PhysRevE93_50101}
as well as the effect due to quantum coherence and correlations \citep{Scully_2003,Jahnke2007,Karimi_2016,Perarnau-Llobet2015PhysRevX5_41011,Su2018}.
The key motivation is to optimize heat engine by improving efficiency
while maintaining the output power. Recently, significant effort has
been devoted to optimizing the Carnot-like heat engine \citep{Esposito_2010,Shiraishi2016PhysRevLett117_190601,Ryabov2016PhysRevE93_50101,Ma2018PhysRevE98_42112,Scopa2018},
similar to the Carnot cycle yet with finite operation time. The price
to pay for such finite-time cycle is the irreversible entropy production,
which was found to be inversely proportional to the control time in
the isothermal process. With this relation, the trade-off between
efficiency and power is explicitly expressed by the constraint formula
derived with different approaches \citep{Broeck2005PhysRevLett95_190602,Tu2008JPhysAMathTheor41_312003,Esposito_2010,Whitney2014PhysRevLett112_130601,Shiraishi2016PhysRevLett117_190601,Ryabov2016PhysRevE93_50101,Holubec2017PhysRevE96_62107,Ma2018PhysRevE98_42112},
along with experimental attempts on the microscopic level \citep{Steeneken_2011,Blickle_2011,Mart_nez_2015,Rossnagel_2016}.

Designing optimal heat engine with Carnot-like engine is a straightforward
approach noticing the Carnot bound is achieved by, yet should not
be limited to. In the theoretical investigation, heat engines with
finite-time Otto cycles hint good performance \citep{Campisi2016NatCommun7_,1812.05089,Karimi_2016},
by utilizing the properties of phase transitions \citep{Campisi2016NatCommun7_,Ma2017PhysRevE96_22143}
or the specific control schemes \citep{Chen2010PhysRevLett104_63002,Torrontegui201362_117,Abah2017EPL118_40005,Deng2018SciAdv4_5909}.
However, the optimization of the finite-time Otto-like heat engine
remains vague, though with many pioneering investigations with concrete
models \citep{Abah2012,Rosnagel2014,Campisi2016NatCommun7_,1812.05089},
mainly due to the difficulty to include the effect of finite-time
operations, especially the finite-time adiabatic processes. The evaluation
of the finite-time effect of adiabatic process is the key to the optimization
of the Otto cycle as well as the Carnot-like cycle.

In this paper, we overcome the current obstacle in optimizing the
quantum Otto cycle by utilizing the quantum adiabatic approximation
\citep{Sun_1988,Wilczek1989_,Rigolin_2008}. In Sec. \ref{sec: 1/t^2 scaling},
we show the universal $\mathcal{C}/\tau^{2}$ scaling of the extra
work during the adiabatic process with long control time $\tau$,
similarly to the scaling of the irreversible entropy production in
finite-time isothermal processes. The impact of the control scheme
is reflected in the coefficient $\mathcal{C}$ via non-adiabatic transitions
between quantum states. In Sec. \ref{sec:Efficiency-at-maximum power},
we optimize the output power of the finite-time quantum Otto engine
based on the $\mathcal{C}/\tau^{2}$ scaling. The efficiency at maximum
power is found in an analytical form, which is probable to exceed
that of the Carnot-like engine. In Sec. \ref{sec:piston model}, the
current formalism is applied to the piston model, which can be solved
analytically to validate the $\mathcal{C}/\tau^{2}$ scaling. The
conclusion is given in Sec. \ref{sec:Conclusion}.

\section{$\mathcal{C}/\tau^{2}$\textit{ }scaling of extra work in the finite-time
adiabatic process\label{sec: 1/t^2 scaling}}

The Otto cycle consists of two adiabatic and two isochoric processes.
The work is performed in the two adiabatic processes, via changing
the controllable parameters $\vec{R}\left(t\right),$ e. g. the volume
for the trapped gas. The time for the isochoric process is typically
negligible comparing to that of the adiabatic process \citep{ChotorlishviliPRE2016,AbahPRE2018}.
We thus focus on the finite-time quantum dynamics during the adiabatic
process.

\begin{figure}
\includegraphics[width=8cm]{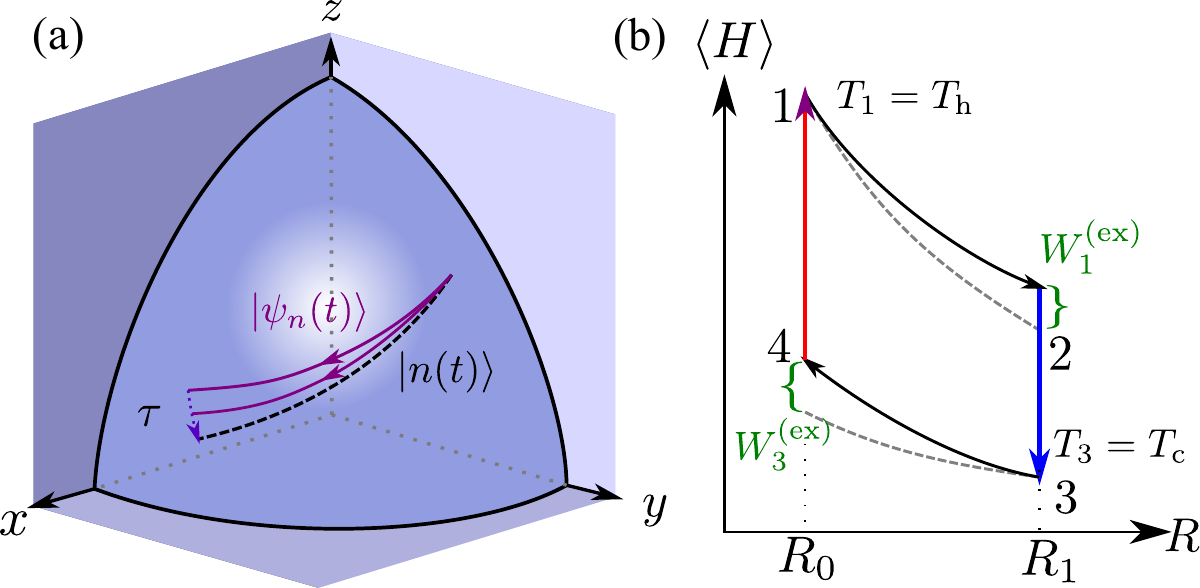}

\caption{(a) The evolution of the quantum state on Bloch sphere during the
adiabatic process. We take the Bloch sphere for the two-level system
as an example. The purple-solid (black-dashed) line presents the finite-time
evolution (the adiabatic trajectory). (b) The energy-control ($\left\langle H\right\rangle -R$)
diagram for the quantum Otto cycle. The solid (dashed) line shows
the finite-time (quasi-static) Otto cycle. To allow the fair comparison,
we set the highest (lowest) temperature $T_{1}$ ($T_{3}$) equal
to that of hot (cold) bath $T_{h}$ ($T_{c}$).}
\end{figure}

At the beginning of adiabatic process, the system is initially prepared
at a thermal equilibrium state 
\begin{equation}
\rho(0)=\frac{e^{-\beta H[\vec{R}(0)]}}{\mathrm{Tr}e^{-\beta H[\vec{R}(0)]}},
\end{equation}
with inverse temperature $\beta$. $H[\vec{R}(t)]$ is the Hamiltonian
with the control parameter $\vec{R}(t)$. The macroscopic parameters
are tuned from $\vec{R}\left(0\right)$ at the beginning to $\vec{R}\left(\tau\right)$
at the end. The evolution of the system during $0<t<\tau$ is controlled
by a time-dependent Hamiltonian $H(t)=H[\vec{R}(t)]$ as 
\begin{equation}
\dot{\rho}=-i[H(t),\rho].
\end{equation}
 Under the instantaneous basis $\left\{ \left|n(t)\right\rangle \right\} $,
the time-dependent Hamiltonian is diagonal 
\begin{equation}
H(t)=\sum_{n}E_{n}(t)\left|n(t)\right\rangle \left\langle n(t)\right|,
\end{equation}
and the initial state is rewritten as $\rho(0)=\sum_{n}p_{n}\left|n(0)\right\rangle \left\langle n(0)\right|$
with the thermal distribution 
\begin{equation}
p_{n}=\frac{e^{-\beta E_{n}(0)}}{\sum_{m}e^{-\beta E_{m}(0)}}.
\end{equation}
 In the adiabatic process, the density matrix at any time in interval
$[0,\tau]$ is 
\begin{equation}
\rho(t)=\sum_{n}p_{n}\left|\psi_{n}(t)\right\rangle \left\langle \psi_{n}(t)\right|,
\end{equation}
 where $\left|\psi_{n}(t)\right\rangle $ follows the Schroedinger
equation 
\begin{equation}
i\frac{\partial}{\partial t}\left|\psi_{n}(t)\right\rangle =H(t)\left|\psi_{n}(t)\right\rangle ,
\end{equation}
with the initial state $\left|\psi_{n}(0)\right\rangle =\left|n(0)\right\rangle $.
The evolution of the instantaneous state $\left|n(t)\right\rangle $
and the state $\left|\psi_{n}(t)\right\rangle $ is illustrated in
Fig. 1(a). The purple-solid lines show the trajectories for finite-time
 adiabatic processes with changing control time $\tau$, and the black-dashed
line presents the evolution of the instantaneous basis. With the increasing
control time $\tau$, the state $\left|\psi_{n}(t)\right\rangle $
approaches to the adiabatic trajectory $\left|n(t)\right\rangle $.

The finite-time effect is reflected through the work extraction during
the adiabatic process. Since the system is isolated from any baths
in the adiabatic processes, no heat is generated during the whole
process. The work done equals to the change of the internal energy
$W(\tau)=\mathrm{Tr}[\rho(\tau)H(\tau)-\rho(0)H(0)]$, explicitly
as 
\begin{equation}
W(\tau)=\sum_{n}p_{n}[\langle\psi_{n}\left(\tau\right)|H\left(\tau\right)|\psi_{n}\left(\tau\right)\rangle-E_{n}\left(0\right)].
\end{equation}
 To show the difference between the finite-time adiabatic process
and its quasi-static counterpart, we define the extra work as 
\begin{equation}
W^{(\mathrm{ex})}(\tau)=W(\tau)-W_{\mathrm{adi}},\label{eq:wexwork}
\end{equation}
where $W_{\mathrm{adi}}=\sum_{n}p_{n}[E_{n}(\tau)-E_{n}(0)]$ is the
work done during the quasi-static adiabatic process. One property
of the extra work for the finite-time adiabatic process is its non-negativity
$W^{(\mathrm{ex})}(\tau)\geq0$, which is proved with details in Appendix
\ref{sec:Thepositivityof}. The non-negativity of the extra work ensures
a lower efficiency of the finite-time Otto cycle than that of the
quasi-static one. Such non-negativity was previously known as the
minimal work principle: For an initial thermal state, the quasi-static
adiabatic process generates the minimal work when the energy level
does not cross \citep{Allahverdyan_2005}.

The key is to obtain the extra work via the dynamics of wave-function
$\left|\psi_{n}(t)\right\rangle $, which is expanded in the instantaneous
basis $\{|l(t)\rangle\}$ as 
\begin{equation}
\left|\psi_{n}(t)\right\rangle =\sum_{l}c_{nl}(t)e^{-i\phi_{l}\left(t\right)}\left|l(t)\right\rangle ,
\end{equation}
 where $\phi_{l}\left(t\right)=\int_{0}^{t}E_{l}(t^{\prime})dt^{\prime}$
is the dynamical phase. The amplitude $c_{nl}(t)$ is obtained by
using the adiabatic perturbation theory \citep{Sun_1988,Rigolin_2008},
where $\nu=1/\tau$ is treated as the perturbation parameter for long
operation time. Based on the high-order adiabatic approximation \citep{Sun_1988},
we obtain $c_{nl}(\tau)$ to the first order in Appendix \ref{sec:HigherorderAdiabaticapproxima}
as\begin{widetext}
\begin{equation}
c_{nl}^{[1]}(\tau)=\begin{cases}
e^{i\tau\tilde{\gamma}_{n}\left(1\right)} & n=l\\
-i\nu[\tilde{T}_{nl}\left(1\right)e^{-i\tau(\tilde{\phi}_{n}\left(1\right)-\tilde{\phi}_{l}\left(1\right))+i\tilde{\gamma}_{n}\left(1\right)}-\tilde{T}_{nl}\left(0\right)e^{i\tilde{\gamma}_{l}\left(1\right)}] & n\neq l
\end{cases},\label{eq:APTamplitude}
\end{equation}
\end{widetext}where the function with tilde is rewritten with the
rescaled time parameter $s=t/\tau$. Here, the Berry phase is given
by
\begin{equation}
\tilde{\gamma}_{l}\left(s\right)=i\int_{0}^{s}\tilde{\Gamma}_{ll}(s^{\prime})ds^{\prime},
\end{equation}
 with the notation $\tilde{\Gamma}_{lm}(s)=\left\langle \tilde{l}(s)\right|d/ds\left|\tilde{m}(s)\right\rangle $;
The non-adiabatic transition rate is
\begin{equation}
\tilde{T}_{nl}\left(s\right)=\frac{\tilde{\Gamma}_{ln}(s)}{\tilde{E}_{n}(s)-\tilde{E}_{l}(s)},
\end{equation}
presenting the transition from the state $\left|\tilde{n}(s)\right\rangle $
with the eigen-energy $\tilde{E}_{n}\left(s\right)$ to another state
$|\tilde{l}(s)\rangle$. For the quasi-static adiabatic process $\nu\rightarrow0$,
the first order term vanishes and the state $\left|\psi_{n}(t)\right\rangle $
remains on the instantaneous eigen-state $\left|n(t)\right\rangle $
with a time-dependent phase. In turn, our definition of the extra
work is appropriate in the sense of retaining quantum adiabatic limit
\citep{Quan_2007}. Here, we clarify the distinction and connection
between quantum adiabacity and thermodynamic adiabacity. Quantum adiabacity
means that population of the energy eigen-states remains unchanged
during the whole process, while thermodynamic adiabacity indicates
no heat exchange between the system and the environment. In the finite-time
adiabatic processes, the unitary evolution of an isolated quantum
system ensures the thermodynamic adiabacity, but quantum adiabacity
is not usually satisfied due to the non-adiabatic transition between
different eigen-states. The rigorous quantum adiabaticity only holds
at the infinite control time limit $\nu\rightarrow0$. The quantum
non-adiabaticity is responsible for the extra work needed to complete
the adiabatic process in finite time.

With the amplitude $c_{nl}(\tau)$, the extra work by Eq. (\ref{eq:wexwork})
is simplified as 
\begin{equation}
W^{\mathrm{(ex)}}(\tau)=\sum_{n,l\neq n}p_{n}[\tilde{E}_{l}(1)-\tilde{E}_{n}(1)]|c_{nl}(\tau)|^{2}.\label{eq:extraworkwithc}
\end{equation}
From the first-order adiabatic approximation result by Eq. (\ref{eq:APTamplitude}),
the value of the absolute square $\left|c_{nl}^{[1]}(\tau)\right|^{2},\,n\ne l$
is divided into the mean part and the oscillating part as
\begin{equation}
\left|c_{nl}^{[1]}(\tau)\right|^{2}=\left(\left|c_{nl}^{[1]}(\tau)\right|^{2}\right)^{\mathrm{(mean)}}+\left(\left|c_{nl}^{[1]}(\tau)\right|^{2}\right)^{\mathrm{(osc)}},
\end{equation}
where the mean part is 
\begin{equation}
\left(\left|c_{nl}^{[1]}(\tau)\right|^{2}\right)^{\mathrm{(mean)}}=\frac{1}{\tau^{2}}\left(\left|\tilde{T}_{nl}\left(1\right)\right|^{2}+\left|\tilde{T}_{nl}\left(0\right)\right|^{2}\right),\label{eq:cmean}
\end{equation}
and the oscillating part is
\begin{widetext}
\begin{equation}
\left(\left|c_{nl}^{[1]}(\tau)\right|^{2}\right)^{\mathrm{(osc)}}=-\frac{2}{\tau^{2}}\mathrm{Re}\left[e^{-i\tau[\tilde{\phi}_{n}(1)-\tilde{\phi}_{l}(1)]+i[\tilde{\gamma}_{n}(1)-\tilde{\gamma}_{l}(1)]}\tilde{T}_{nl}\left(1\right)\tilde{T}_{nl}^{*}\left(0\right)\right].\label{eq:cosc}
\end{equation}
\end{widetext}

To the first order, $|c_{nl}^{[1]}(\tau)|^{2}$ is proportional to
$\nu^{2}$, leading to the $\mathcal{C}/\tau^{2}$ scaling of the
extra work. This scaling is different from the $1/\tau$ scaling of
the irreversible entropy production in the finite-time isothermal
process of the Carnot-like cycle \citep{Shiraishi2016PhysRevLett117_190601,Ma2018PhysRevE98_42112}.
Corresponding to Eqs. (\ref{eq:cmean}) and (\ref{eq:cosc}), the
extra work $W^{(\mathrm{ex})}(\tau)=W^{(\mathrm{mean})}(\tau)+W^{(\mathrm{osc})}(\tau)$
by Eq. (\ref{eq:extraworkwithc}) is divided into the mean extra work
\begin{align}
W^{(\mathrm{mean})}(\tau) & =\frac{\Sigma}{\tau^{2}},\label{meanwork}
\end{align}
and the oscillating extra work

\begin{align}
W^{(\mathrm{osc})}(\tau) & =\frac{\omega\left(\tau\right)}{\tau^{2}}.\label{osiwork}
\end{align}
The mean extra work decreases decreases monotonously with the increasing
control time $\tau$, while the oscillating extra work oscillates
around $0$, and contributes the fluctuation in the extra work. The
coefficients in Eqs. (\ref{meanwork}) and (\ref{osiwork}) follow
explicitly as 
\begin{equation}
\Sigma=\sum_{n,l\ne n}p_{n}[\tilde{E}_{l}(1)-\tilde{E}_{n}(1)][|\tilde{T}_{nl}\left(1\right)|^{2}+|\tilde{T}_{nl}\left(0\right)|^{2}],\label{eq:meancoe}
\end{equation}
and
\begin{widetext}
\begin{equation}
\omega(\tau)=-\sum_{n,l\ne n}2p_{n}[\tilde{E}_{l}(1)-\tilde{E}_{n}(1)]\mathrm{Re}\{\tilde{T}_{nl}\left(1\right)\tilde{T}_{nl}^{*}\left(0\right)e^{-i\tau(\tilde{\phi}_{n}\left(1\right)-\tilde{\phi}_{l}\left(1\right))+i(\tilde{\gamma}_{n}\left(1\right)-\tilde{\gamma}_{l}\left(1\right))}\}.\label{eq:osccoe}
\end{equation}
\end{widetext}

The impact of the control scheme is reflected through the transition
amplitude $\tilde{T}_{nl}\left(s\right)$. Interestingly, the mean
extra work, to the leading order, only depends on the initial (final)
transition amplitude $\tilde{T}_{nl}\left(0\right)$ ($\tilde{T}_{nl}\left(1\right)$),
instead of the whole trajectory. And the oscillating one relies on
the trajectory only through the dynamical phase $\tilde{\phi}_{n}\left(s\right)$
and the Berry phase $\tilde{\gamma}_{n}\left(s\right)$.

For the oscillating extra work, $\omega(\tau)$ oscillates around
$0$ with the increasing control time $\tau$. When we consider the
system with the incommensurable energy difference $\tilde{E}_{l}(s)-\tilde{E}_{n}(s)$
for different sets of indexes $l$ and $n$, the oscillation of $\omega(\tau)$
contains different frequency $\tilde{\phi}_{n}\left(1\right)-\tilde{\phi}_{l}\left(1\right)$.
In the summation of $\omega\left(\tau\right)$, the terms with different
phase $\tilde{\phi}_{n}\left(1\right)-\tilde{\phi}_{l}\left(1\right)$
cancel out each other. In the follow discussion, we will neglect the
oscillating term in Eq. (\ref{osiwork}). Yet, this oscillating terms
may introduce higher efficiency for system with few energy levels,
e.g. the two-level system \citep{1812.05089}.

\section{Efficiency at maximum power for quantum Otto heat engine\label{sec:Efficiency-at-maximum power}}

With the $\mathcal{C}/\tau^{2}$ scaling of the extra work, we evaluate
the performance of quantum Otto engine by the efficiency and the output
power. The Otto cycle is illustrated via the $\left\langle H\right\rangle -\vec{R}$
diagram in Fig. 1(b). The solid line shows the finite-time Otto cycle,
while the dashed line shows the corresponding quasi-static one. The
work done during the two adiabatic processes ($1\rightarrow2$) and
($3\rightarrow4$) are $W_{1}\left(\tau_{1}\right)<0$ and $W_{3}(\tau_{3})>0$
with the change of external parameters ($R_{0}\leftrightarrow R_{1}$)
respectively. The heat engine contacts with the hot (cold) bath and
reaches the equilibrium with the temperature $T_{1}$ ($T_{3}$) in
the isochoric heating ($4\rightarrow1$) (isochoric cooling $(2\rightarrow3)$)
with the fixed parameter $R_{0}$ ($R_{1}$).

The performance of quantum Otto engine is evaluated by the efficiency
and the output power. We need the net work and the heat exchange under
the adiabatic perturbation approximation. For the two adiabatic processes,
the work is 
\begin{align}
W_{i}\left(\tau_{i}\right) & \equiv\mathrm{Tr}[\rho_{i+1}H_{i+1}]-\mathrm{Tr}[\rho_{i}H_{i}]\\
 & =W_{i}^{\mathrm{adi}}+\frac{\Sigma_{i}}{\tau_{i}^{2}},\:i=1,3,
\end{align}
where $\tau_{i}$ is the corresponding control time, and $\Sigma_{i}$
is the corresponding coefficients related to the control scheme. The
work in the quasi-static adiabatic process is given by
\begin{equation}
W_{i}^{\mathrm{adi}}=\mathrm{Tr}[\rho_{i+1}^{\mathrm{adi}}H_{i+1}]-\mathrm{Tr}[\rho_{i}H_{i}].
\end{equation}
 We consider the relaxation time in the isochoric processes is much
shorter than the control time $\tau_{i},\,i=1,3$ in the adiabatic
processes. The time consuming of the isochoric processes is neglected,
and the system is fully thermalized after the isochoric processes.
The heat exchange with the hot bath during the isochoric process is
\begin{align}
Q_{h} & \equiv\mathrm{Tr}[\rho_{4}H_{4}]-\mathrm{Tr}[\rho_{1}H_{1}]\\
 & =Q_{h}^{\mathrm{adi}}-\frac{\Sigma_{3}}{\tau_{3}^{2}}.
\end{align}
 The net work for the whole cycle is 
\begin{equation}
W_{\mathrm{T}}=-[W_{1}\left(\tau_{1}\right)+W_{3}\left(\tau_{3}\right)],
\end{equation}
 and the efficiency is 
\begin{equation}
\eta=\frac{W_{\mathrm{T}}}{Q_{h}}.
\end{equation}
 Combing the equations for $W_{i}(\tau_{i})$ and $W_{i}^{\mathrm{adi}}$,
the power $P=W_{\mathrm{T}}/(\tau_{1}+\tau_{3})$ for the finite-time
Otto heat engine follows explicitly 
\begin{align}
P & =\frac{W_{\mathrm{T}}^{\mathrm{adi}}}{\tau_{1}+\tau_{3}}-\frac{1}{\tau_{1}+\tau_{3}}\left(\frac{\Sigma_{1}}{\tau_{1}^{2}}+\frac{\Sigma_{3}}{\tau_{3}^{2}}\right),\label{eq:power}
\end{align}
with the efficiency

\begin{align}
\eta & =\frac{W_{\mathrm{T}}^{\mathrm{adi}}-\left(\Sigma_{1}/\tau_{1}^{2}+\Sigma_{3}/\tau_{3}^{2}\right)}{W_{\mathrm{T}}^{\mathrm{adi}}/\eta^{\mathrm{adi}}-\Sigma_{3}/\tau_{3}^{2}}.\label{eq:eff}
\end{align}
Here, $W_{\mathrm{T}}^{\mathrm{adi}}=-(W_{1}^{\mathrm{adi}}+W_{3}^{\mathrm{adi}})$
is the net work for the quasi-static Otto cycle with the corresponding
efficiency $\eta^{\mathrm{adi}}=W_{\mathrm{T}}^{\mathrm{adi}}/Q_{h}^{\mathrm{adi}}$.

According to the optimal condition of the maximum power $\partial P/\partial\tau_{1}=0,\,\partial P/\partial\tau_{3}=0$,
the current finite-time Otto engine reaches its maximum power 
\begin{equation}
P_{\mathrm{max}}=2\left[\frac{W_{\mathrm{T}}^{\mathrm{adi}}}{3\left(\Sigma_{1}^{1/3}+\Sigma_{3}^{1/3}\right)}\right]^{\frac{3}{2}}\label{eq:maxpower}
\end{equation}
at the optimal operation time $\tau_{1}^{*}=[3(\Sigma_{1}^{2/3}\Sigma_{3}^{1/3}+\Sigma_{1})/W_{\mathrm{T}}^{\mathrm{adi}}]^{1/2}$
and $\tau_{3}^{*}=[3(\Sigma_{3}^{2/3}\Sigma_{1}^{1/3}+\Sigma_{3})/W_{\mathrm{T}}^{\mathrm{adi}}]^{1/2}$.
The corresponding efficiency at the maximum power (EMP) is 
\begin{align}
\eta_{\mathrm{EMP}} & =\frac{2\eta^{\mathrm{adi}}}{3-\frac{\eta^{\mathrm{adi}}}{1+(\Sigma_{1}/\Sigma_{3})^{1/3}}},\label{Emp}
\end{align}
which depends on the ratio $\Sigma_{1}/\Sigma_{3}$, and the efficiency
$\eta^{\mathrm{adi}}$ of the quasi-static Otto cycle. In the limit
$\Sigma_{1}/\Sigma_{3}\rightarrow0$, the EMP reaches the upper bound
\begin{equation}
\eta_{\mathrm{EMP}}^{+}=\frac{2\eta^{\mathrm{adi}}}{3-\eta^{\mathrm{adi}}}.
\end{equation}
 In the limit $\Sigma_{1}/\Sigma_{3}\rightarrow\infty$, the EMP reaches
the lower bound 
\begin{equation}
\eta_{\mathrm{EMP}}^{-}=\frac{2}{3}\eta^{\mathrm{adi}}.
\end{equation}

We obtain the main result in Eq. (\ref{Emp}) with the first-order
quantum adiabatic approximation, where the inverse control time $\nu$
is the perturbation parameter. The result relies on two key factors,
i.e, the long control time \citep{Sun_1988,Rigolin_2008} $\tau$
and the non-level crossing condition \citep{Allahverdyan_2005}. To
obtain the EMP, we have neglected the oscillating  extra work with
the observation of incommensurability of the typical energy levels.
Yet, such oscillating part can introduce interesting effects on EMP
for small quantum systems, e.g., the minimal quantum heat engine with
two-level system \citep{Linden2010PhysRevLett105_130401}.

\begin{figure}
\includegraphics[width=8cm]{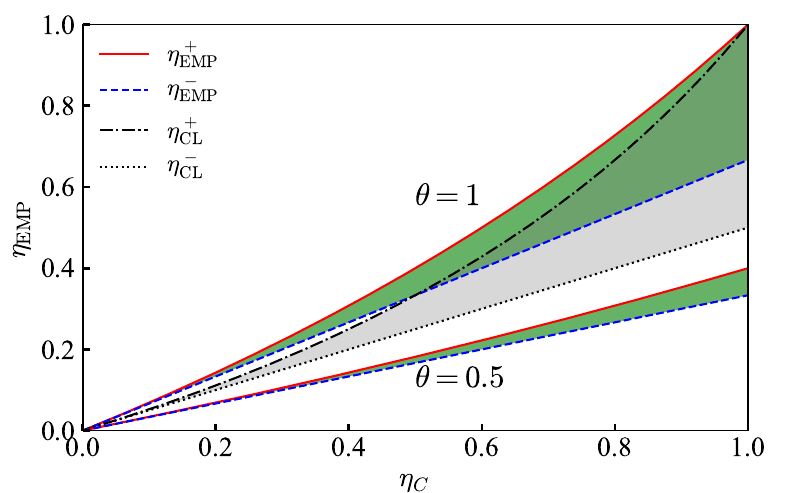}

\caption{Comparison between the EMP of finite-time quantum Otto cycle and Carnot-like
cycle. The red-solid (blue-dashed) lines show the upper (lower) for
the EMP of finite-time Otto cycle with two sets of $\theta$. The
black-dashdotted (black-dotted) line presents the upper (lower) bound
$\eta_{\mathrm{CL}}^{+}$ ( $\eta_{\mathrm{CL}}^{-}$) for the EMP
of Carnot-like cycle.}
\label{fig2}
\end{figure}

We turn to compare the EMP of the finite-time Otto cycle with that
of the Carnot-like cycle. For the Carnot-like cycle, the upper bound
\citep{Esposito_2010,Ma2018PhysRevE98_42112} is $\eta_{\mathrm{CL}}^{+}=\eta_{\mathrm{C}}/(2-\eta_{\mathrm{C}})$,
where $\eta_{\mathrm{C}}=1-T_{c}/T_{h}$ is the Carnot efficiency
for heat engine working between the low temperature $T_{c}$ and the
high temperature $T_{h}$ baths. To allow a fair comparison, we set
the highest (lowest) temperature $T_{1}$ ($T_{3}$) in the isochoric
process to be the temperature for the hot (cold) bath, namely $T_{1}=T_{h}$
($T_{3}=T_{c}$). To surpass the EMP of Carnot-like heat engine ($\eta_{\mathrm{EMP}}^{+}>\eta_{\mathrm{CL}}^{+}$),
it is required that 
\begin{equation}
\eta^{\mathrm{adi}}>\frac{3\eta_{\mathrm{C}}}{4-\eta_{\mathrm{C}}}.
\end{equation}
 The efficiency $\eta^{\mathrm{adi}}$ of the quasi-static Otto is
always smaller than the Carnot efficiency $\eta_{\mathrm{C}}$, namely,
$\eta^{\mathrm{adi}}<\eta_{\mathrm{C}}$.

In Fig. 2, we show the EMP of both the Carnot-like cycle and the current
quantum Otto cycle. We set the efficiency of the corresponding quasi-static
Otto heat engine as $\eta^{\mathrm{adi}}=\theta\eta_{\mathrm{C}}$
with the ratio $\theta\in[0,1]$. For the finite-time quantum Otto
cycle, the upper (lower) bound $\eta_{\mathrm{EMP}}^{+}$ ($\eta_{\mathrm{EMP}}^{-}$)
is plotted as the red-solid (blue-dashed) line. Two sets of the ratios
$\theta=0.5$ and $\theta=1$ are plotted. For the Carnot-like heat
engine, the black-dashdotted and the black-dotted lines give the upper
bound $\eta_{\mathrm{CL}}^{+}$ and the lower bound $\eta_{\mathrm{CL}}^{-}=\eta_{C}/2$
respectively \citep{Esposito_2010}. For $\theta=1$, the curve shows
that the EMP of the finite-time quantum Otto cycle exceeds the one
for the finite-time Carnot cycle. Such higher EMP is achievable only
at the region $\theta>3/4$. The curves for $\theta=0.5$ show the
lower efficiency than that of the Carnot-like cycle. The current generic
model implies the possibility to surpass the EMP of the Carnot-like
cycle by choosing the proper efficiency of the quasi-static Otto cycle
$\eta^{\mathrm{adi}}$ larger than $3\eta_{C}/4$. We will realize
such Otto cycle with an example of the quantum piston model.

\section{Finite-time quantum Otto engine on piston model\label{sec:piston model}}

We illustrate the $\mathcal{C}/\tau^{2}$ scaling of extra work during
the adiabatic process with the widely used quantum piston model \citep{Makowski_1991,Doescher_1969,Quan2012}
and show the surpassed EMP of Carnot-like engines with the designed
finite-time Otto cycle. Now we consider a concrete model of a single
particle trapped in a square box with the Hamiltonian,
\begin{align}
H(t) & =-\frac{1}{2M}\frac{\partial^{2}}{\partial x^{2}}+V(x,t).\label{eq:timedependenthamiltonian}
\end{align}
where $M$ is the mass of the particle, and $V(x,t)$ is the square
potential 
\begin{equation}
V(x,t)=\begin{cases}
\infty & x<0,\,x>L(t)\\
0 & 0\leq x\leq L(t)
\end{cases}.
\end{equation}

The controllable length $L(t)$ serves as the tuning parameter $R\left(t\right)$
as discussed in the generic model. The advantage of the current model
is the existence of an exact solution for the linear control protocol
\citep{Makowski_1991,Doescher_1969}
\begin{equation}
L(t)=L_{0}+(L_{1}-L_{0})\frac{t}{\tau},
\end{equation}
 which allows a direct validation of the scaling in Eq. (\ref{meanwork})
derived by the adiabatic perturbation theory. Here, $L_{0}$ ($L_{1}$)
is the initial (final) length of the box during the adiabatic process.

For this control scheme $\tilde{L}\left(s\right)=L\left(s\tau\right)$,
the instantaneous wave-function is 
\begin{equation}
\left\langle x\right.\left|\tilde{n}\left(s\right)\right\rangle =\sqrt{\frac{2}{\tilde{L}}}\sin\left(\frac{n\pi}{\tilde{L}}x\right),\label{eq:instantaneouswavefunction}
\end{equation}
 with the corresponding energy 
\begin{equation}
\tilde{E}_{n}\left(s\right)=\frac{\pi^{2}n^{2}}{2M\tilde{L}^{2}}.\label{eq:correspondingenergy}
\end{equation}
 The non-adiabatic transition rate is 
\begin{align}
\tilde{T}_{nl}(s) & =-\frac{4Mnl(-1)^{l+n}(L_{1}-L_{0})}{\pi^{2}(n^{2}-l^{2})^{2}}\tilde{L}(s).
\end{align}
Under long control time limit, we obtain the asymptotic result of
the extra work as 
\begin{align}
W^{\mathrm{(mean)}}(\tau) & =\frac{ML_{1}^{2}(1-r)^{2}(1+r^{2})}{\tau^{2}}(\frac{1}{6}-\sum_{n=1}^{\infty}\frac{p_{n}}{4\pi^{2}n^{2}}),\label{eq:91}
\end{align}
with the expansion ratio $r=L_{0}/L_{1}$. The initial thermal distribution
is 
\begin{equation}
p_{n}(\beta,L_{0})=\frac{e^{-\frac{\beta\pi^{2}n^{2}}{2L_{0}^{2}M}}}{Z(\beta,L_{0})},\label{eq:pistoninitialthermal distribution}
\end{equation}
with the initial inverse temperature $\beta=1/k_{B}T$. The partition
function is 
\begin{equation}
Z(\beta,L)=\frac{1}{2}\vartheta_{3}\left(0,e^{-\frac{\beta\pi^{2}}{2L^{2}M}}\right)-\frac{1}{2},
\end{equation}
 where $\vartheta_{3}\left(0,q\right)=2\sum_{n=1}^{\infty}q^{n^{2}}+1$
is the Elliptic-Theta function. The detailed derivation of Eq. (\ref{eq:91})
is given in Appendix \ref{sec:1DPistonModel}, where the oscillating
extra work is also obtained analytically. At high temperature limit
$\beta\rightarrow0$, the thermal de Broglie wavelength $\lambda_{\mathrm{th}}=(2\pi\beta/M)^{1/2}$
is much smaller than the length of the box and the summation $\sum_{n=1}^{\infty}p_{n}/(4\pi^{2}n^{2})$
in Eq. (\ref{eq:91}) can be neglected. Thus, we obtain the approximation
for the mean extra work in Eq. (\ref{eq:91}) as 
\begin{equation}
W^{(\mathrm{mean})}(\tau)\approx\frac{ML_{1}^{2}(1-r)^{2}(1+r^{2})}{6\tau^{2}}.\label{eq:wmeanpistonappro}
\end{equation}

\begin{figure}
\includegraphics[width=8cm]{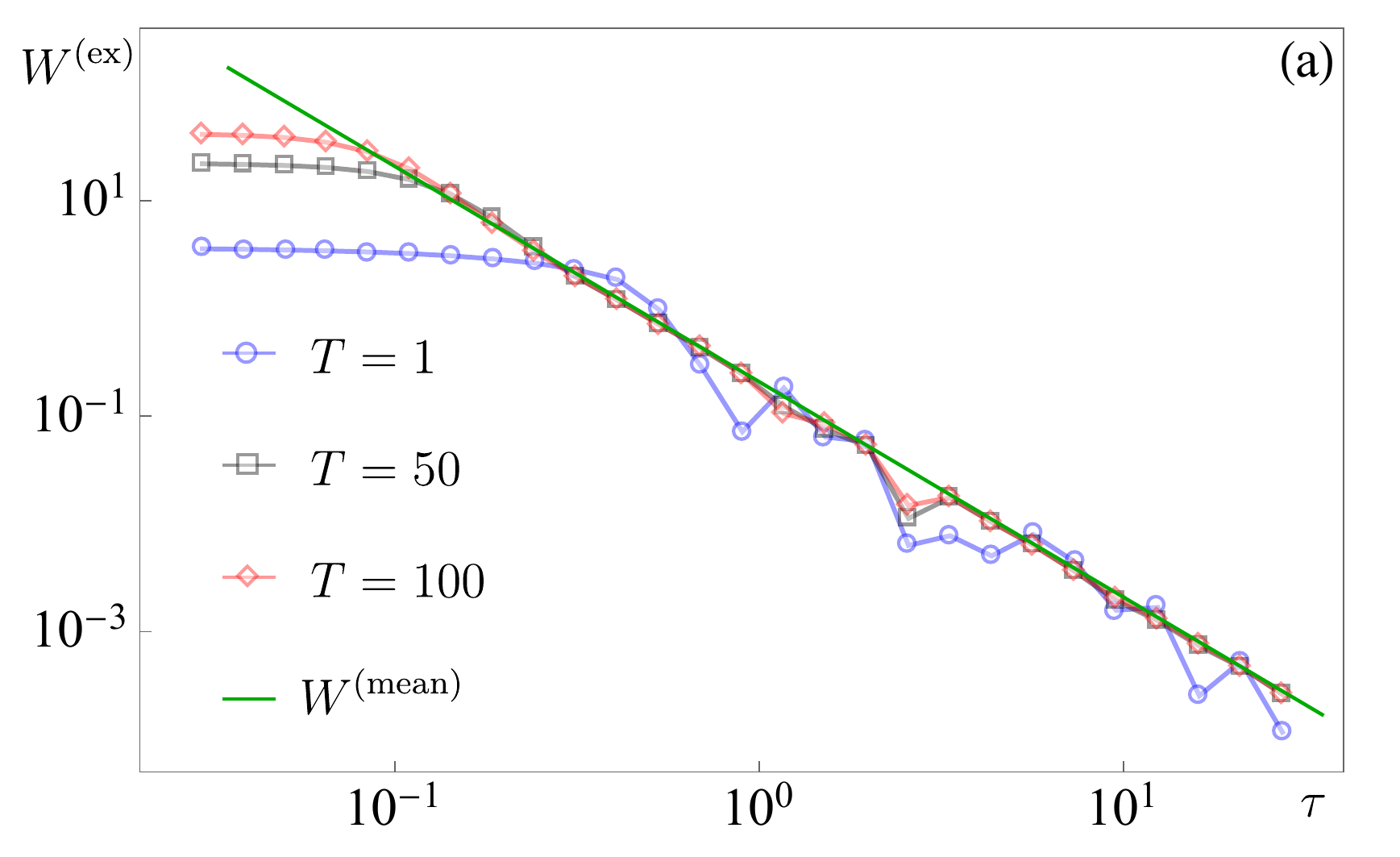}

\includegraphics[width=8cm]{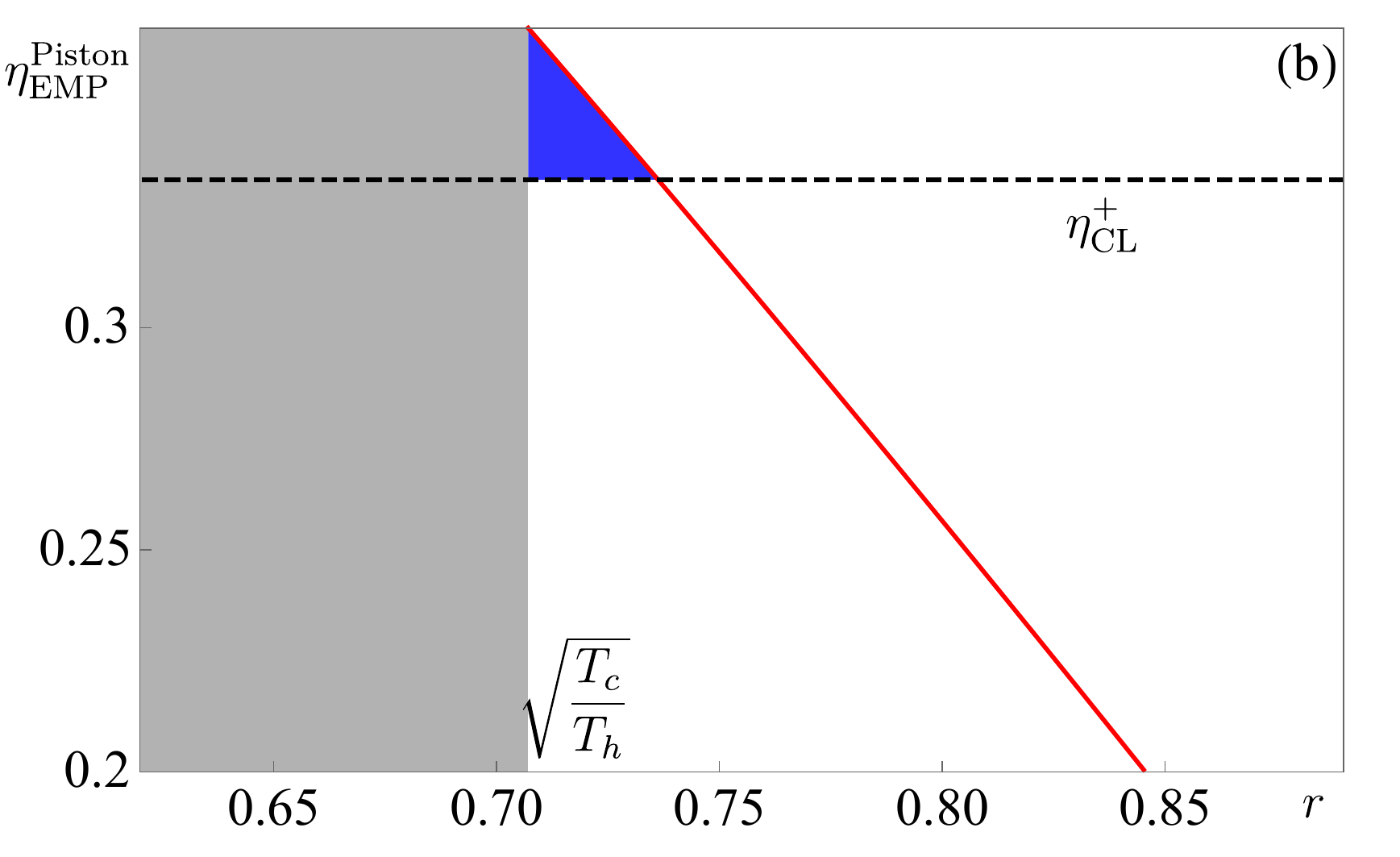}

\caption{(a) The $\mathcal{C}/\tau^{2}$ scaling for the extra work in the
finite-time adiabatic expansion process on quantum piston model, with
the length chosen as $L_{0}=1$ and $L_{1}=2$. The blue circle, black
square, and red diamaond show the exact numerical result with the
initial temperatures $T=1,\,50,\,100$, respectively. The green-solid
line presents the analytical result of the mean extra work for large
$\tau$ in Eq. (\ref{meanwork}). (b) The EMP of the finite-time quantum
Otto cycle as the function of the expansion ratio $r=L_{0}/L_{1}$
(the red-solid line), with the fixed temperature ratio $T_{c}/T_{h}=1/2$.
The black-dashed horizontal line presents the upper bound of the Carnot-like
cycle $\eta_{\mathrm{CL}}^{+}$. The gray area shows the engine with
negative output power $P<0$, and the blue area shows the higher EMP
than that of the Carnot-like cycle.}
\label{fig3}
\end{figure}

By controlling the length of the trap, we realize the finite-time
quantum Otto cycle with the current quantum piston model. The two
lengths for the adiabatic process are $L_{0}$ and $L_{1}$ with $L_{0}<L_{1}$.
For the quasi-static Otto cycle, the efficiency of the engine is $\eta^{\mathrm{adi}}=1-r^{2}$
and the net work of the whole cycle has a simple result at high temperature
\begin{equation}
W^{\mathrm{adi}}=\frac{k_{B}}{2}\left(T_{h}r^{2}-T_{c}\right)(\frac{1}{r^{2}}-1).
\end{equation}
 For the finite-time adiabatic process, the coefficients of the mean
extra work with linear control schemes are 
\begin{equation}
\Sigma_{1}=\frac{M(1-r)^{2}\left(1+r^{2}\right)L_{1}^{2}}{6},
\end{equation}
 and 
\begin{equation}
\Sigma_{3}=\frac{M(1-r)^{2}\left(1+r^{2}\right)L_{1}^{2}}{6r^{2}}.
\end{equation}

Fig. \ref{fig3}(a) validates the $\mathcal{C}/\tau^{2}$ scaling
of the extra work $W^{(\mathrm{mean})}(\tau)$ in Eq. (\ref{eq:wmeanpistonappro})
during the expansion of the quantum piston model. We set the mass
$M=1$ and the Boltzmann constant as $k_{B}=1$ in all the later calculation.
During the expansion, the length of the box varies from the initial
value $L_{0}=1$ to the final value $L_{1}=2$. Exact results are
obtained with analytical solution of the time-dependent Schroedinger
equation in Ref. \citep{Doescher_1969,Makowski_1991,Quan2012}. We
choose the initial thermal states with different temperatures $T=1,$
$50$ and $100$, marked with blue circle, black square, and red diamond
respectively. The oscillation of the extra work becomes weaker for
higher temperature. For long control time, the exact numerical result
of the extra work (the markers) matches with the analytical one (the
green-solid line), demonstrating the $\mathcal{C}/\tau^{2}$ scaling
of the extra work.

The maximum power for the piston model is obtained as 
\begin{align}
P_{\mathrm{max}}^{\mathrm{Piston}} & =\frac{1}{3L_{1}}\left(\frac{k_{B}(T_{h}r^{2}-T_{c})(1-r^{2})}{[M(1-r)^{2}\left(1+r^{2}\right)]^{\frac{1}{3}}(r^{2}+r^{\frac{4}{3}})}\right)^{\frac{3}{2}},\label{eq:43}
\end{align}
by choosing the optimal control time $\tau_{1}^{*}$ and $\tau_{3}^{*}$.
And the corresponding efficiency is obtained by Eq. (\ref{Emp})
\begin{equation}
\eta_{\mathrm{EMP}}^{\mathrm{Piston}}=\frac{2\eta^{\mathrm{adi}}}{3-\frac{\eta^{\mathrm{adi}}}{(1-\eta^{\mathrm{adi}})^{1/3}+1}}.\label{eq:emppiston}
\end{equation}
The detailed derivation of Eqs. (\ref{eq:43}) and (\ref{eq:emppiston})
together with the optimal control time $\tau_{1}^{*}$ and $\tau_{3}^{*}$
is given in Appendix \ref{sec:1DPistonModel}.

In Fig. \ref{fig3}(b), we plot the EMP $\eta_{\mathrm{EMP}}^{\mathrm{Piston}}$
in Eq. (\ref{eq:emppiston}) as the function of the expansion ratio
$r$. The requirement of the positive power in Eq. (\ref{eq:43})
implies the constraint for the expansion ratio $r$ as $\sqrt{T_{c}/T_{h}}<r<1$,
shown as the white area. The upper bound $\eta_{\mathrm{CL}}^{+}$
of the Carnot-like cycle is plotted as the horizontal black-dashed
line, with the fixed temperature ratio $T_{c}/T_{h}=1/2$. The EMP
$\eta_{\mathrm{EMP}}^{\mathrm{Piston}}$ of the current piston model
is shown as the red solid line. Fig. \ref{fig3}(b) indicates the
higher EMP of the finite-time quantum Otto cycle than that of the
Carnot-like cycle at the region $1/\sqrt{2}<r<0.736$, illustrated
as the blue area. The number $0.736$ is obtained by solving the equation
$\eta_{\mathrm{EMP}}^{\mathrm{Piston}}=\eta_{\mathrm{CL}}^{+}$. In
the current model, the piston is controlled with the simplest scheme
that the expansion ratio is the only optimizing parameter for the
EMP $\eta_{\mathrm{EMP}}^{\mathrm{Piston}}$. More complicated control
scheme \citep{Makowski_1991} can be considered to show more flexible
tuning EMP beyond the Carnot-like cycle.

\section{Conclusion \& Remarks\label{sec:Conclusion}}

In this paper, we have studied the finite-time effect of adiabatic
processes. With the high-order adiabatic approximation, we have proved
the universal $\mathcal{C}/\tau^{2}$ scaling for the extra work in
the finite-time adiabatic processes, and validated it with the quantum
piston model. It is meaningful to test this universal scaling on other
complex quantum systems %
from both theoretical and experimental aspects. The current experimental
setup on the trapped Fermi gas \citep{Deng2016,Deng2018SciAdv4_5909}
can be directly applied to verify the $\mathcal{C}/\tau^{2}$ scaling
of the extra work. One needs to choose a fixed protocol of the adiabatic
process, and measures the work to complete the adiabatic process for
different control time $\tau$.

Moreover, we described a new class of finite-time quantum heat engine
with Otto cycle. Importantly, we showed such cycle is capable to achieve
higher efficiency at maximum power than that of the widely-used Carnot-like
cycle. The better performance of the quantum Otto cycle will attract
attentions for new designs of the quantum heat engine, instead of
focusing on optimizing the Carnot-like cycle in finite time. It is
proposed that the quantum Otto cycle can be implemented on a single-ion
engine \citep{Abah2012,Rosnagel2014}. Our study contributes to the
further optimization of the concrete finite-time quantum engine in
the experiments.

In the derivation of the $\mathcal{C}/\tau^{2}$ scaling for the extra
work, the oscillating extra work is neglected due to the incommensurable
energy difference of the complex system. Yet, for the quantum heat
engine with work matter consisting of few energy levels, the oscillating
extra work can affect the performance of the quantum Otto engine.
The oscillating behavior of the extra work leads to a quantum Otto
engine with high efficiency \citep{Peterson2018,Assis2019}, and will
induce new effect on the efficiency-power constraint relation \citep{Chen2019}.
\begin{acknowledgments}
This work is supported by the NSFC (Grants No. 11534002 and No. 11875049),
the NSAF (Grant No. U1730449 and No. U1530401), and the National Basic
Research Program of China (Grants No. 2016YFA0301201 and No. 2014CB921403).
H.D. also thanks The Recruitment Program of Global Youth Experts of
China.
\end{acknowledgments}

\bibliographystyle{apsrev4-1}
\bibliography{finite_time_adiabatic}

\begin{widetext}

\appendix

\section{Positivity of the Extra Work\label{sec:Thepositivityof}}

In this appendix, we prove the positivity of the extra work by using
the Schur-Horn theorem. We remark that the proof of the positive extra
work was already presented elsewhere \citep{Allahverdyan_2005}. However,
our version of the proof with the Schur-Horn theorem is new and straightforward.
Generally, we assume the energy levels shift remaining the order $\tilde{E}_{1}(s)<\tilde{E}_{2}(s)...<\tilde{E}_{n}(s)<...$
during the whole adiabatic process \citep{Allahverdyan_2005}. For
an initial thermal state, the extra work from Eq. (\ref{eq:ex work})
is explicitly written as

\begin{equation}
W^{\mathrm{(ex)}}(\tau)=\sum_{l=1}^{\infty}\lambda_{ll}\tilde{E}_{l}(1)-\sum_{n=1}^{\infty}p_{n}\tilde{E}_{n}(1),\label{Wex}
\end{equation}
with the notation $\lambda_{ll}=\sum_{n=1}^{\infty}p_{n}\left|c_{nl}(\tau)\right|^{2}$.
We rearrange the summation and obtain

\begin{equation}
W^{\mathrm{(ex)}}(\tau)=\tilde{E}_{1}(1)\left[\sum_{l=1}^{\infty}\left(\lambda_{ll}-p_{l}\right)\right]+\sum_{j=2}^{\infty}\left(\tilde{E}_{j}(1)-\tilde{E}_{j-1}(1)\right)\left[\sum_{l=j}^{\infty}\left(\lambda_{ll}-p_{l}\right)\right].\label{Wex-1}
\end{equation}
The first term on the right hand of Eq. (\ref{Wex-1}) is zero due
to the normalized condition for the probability $\sum_{l=1}^{\infty}\lambda_{ll}=\sum_{l=1}^{\infty}p_{l}=1$.
We prove the second term on the right hand side is non-negative based
on Schur-Horn theorem \citep{Horn_1954}. Since $\tilde{E}_{j}(1)-\tilde{E}_{j-1}(1)>0$,
we only need to prove $\sum_{l=j}^{\infty}\lambda_{ll}\geq\sum_{l=j}^{\infty}p_{l}$.

Here, $\lambda_{ll}$ can be regarded as the diagonal element for
the Hermite matrix $\lambda_{lm}=\sum_{n=1}^{\infty}p_{n}\left[c_{nl}(\tau)\right]^{*}c_{nm}(\tau)$,
which is obtained from the diagonal matrix with the diagonal element
$p_{n}$ through the unitary transform $c_{nm}(\tau)$. Thus, the
eigenvalue of this Hermitian matrix is exact $p_{n}$. We re-sequence
the diagonal terms $\lambda_{ll}$ in the non-increasing order as
$\tilde{\lambda}_{11}\geq\tilde{\lambda}_{22}\geq...\geq\tilde{\lambda}_{nn}...$.
Schur-Horn theorem \citep{Horn_1954} presents the following inequality
\begin{equation}
\sum_{l=1}^{j-1}\tilde{\lambda}_{ll}\leq\sum_{l=1}^{j-1}p_{l},\,j\geq2
\end{equation}
for a Hermitian matrix with the diagonal terms $\tilde{\lambda}_{ll}$
and eigenvalue $p_{l}$ both in non-increasing order. Together with
the normalization of the probability, we have the inequality $\sum_{l=j}^{\infty}\tilde{\lambda}_{ll}\geq\sum_{l=j}^{\infty}p_{l},\,j\geq1$.
Since $\tilde{\lambda}_{ll}$ gives the non-increasing order for $\lambda_{ll}$,
we have apparently $\sum_{l=j}^{\infty}\lambda_{ll}\geq\sum_{l=j}^{\infty}\tilde{\lambda}_{ll}$,
and thus $\sum_{l=j}^{\infty}\left(\lambda_{ll}-p_{l}\right)\geq0$.

Therefore, we have proven $W^{\mathrm{(ex)}}(\tau)\geq0$: the extra
work for an initial thermal state is non-negative when the energy
level does not cross during the finite-time adiabatic process.

\section{First-order Adiabatic Approximation and the Extra Work\label{sec:HigherorderAdiabaticapproxima}}

This appendix is devoted to showing the detailed derivation of the
non-adiabatic correction for the extra work for finite-time adiabatic
processes based on higher-order adiabatic approximation \citep{Sun_1988}.
The Schroedinger equation $i\partial_{t}\left|\psi_{n}(t)\right\rangle =H(t)\left|\psi_{n}(t)\right\rangle $
results in the following differential equation for the amplitude $c_{nl}(t)$
\begin{equation}
\frac{d}{dt}c_{nl}(t)+c_{nl}(t)\Gamma_{ll}(t)+\sum_{m\ne l}c_{nm}(t)e^{-i\left(\phi_{m}(t)-\phi_{l}(t)\right)}\Gamma_{lm}(t)=0,
\end{equation}
with the dynamical phase $\phi_{l}(t)=\int_{0}^{t}E_{l}(t^{\prime})dt^{\prime}$
and the notation $\Gamma_{lm}(t)=\left\langle l(t)\right|d/dt\left|m(t)\right\rangle $.
We consider for a given protocol of the adiabatic process $\tilde{H}(s)=H(\tau s)=\sum_{n}\tilde{E}_{n}(s)\left|\tilde{n}(s)\right\rangle \left\langle \tilde{n}(s)\right|$
with $\tau$ as the control time, where $\left|\tilde{n}(s)\right\rangle =\left|n(\tau s)\right\rangle $,
$\tilde{E}_{n}(s)=E_{n}(\tau s)$. Representing the amplitude $b_{nl}(s)=c_{nl}(\tau s)$
with the rescaled time parameter $s$, the differential equation is
rewritten for $b_{nl}(s)$ as
\begin{equation}
\frac{d}{ds}b_{nl}(s)+b_{nl}(s)\tilde{\Gamma}_{ll}(s)+\sum_{m\ne l}b_{nm}(s)e^{-i\tau\left(\tilde{\phi}_{m}(s)-\tilde{\phi}_{l}(s)\right)}\tilde{\Gamma}_{lm}(s)=0,
\end{equation}
where the notation$\tilde{\Gamma}_{lm}(s)=\left\langle \tilde{l}(s)\right|\frac{d}{ds}\left|\tilde{m}(s)\right\rangle $
and the dynamical phase $\tilde{\phi}_{l}(s)=\int_{0}^{s}\tilde{E}_{l}(s^{\prime})ds^{\prime}$
are given by the rescaled time parameter $s$.

Based on the high-order adiabatic approximation in Ref. \citep{Sun_1988},
we obtain the solution of $b_{nl}^{[1]}(s)=b_{nl}^{(0)}(s)+b_{nl}^{(1)}(s)/\tau$
to the first order of $1/\tau$, where $b_{nl}^{(0)}(s)$ and $b_{nl}^{(1)}(s)$
satisfy the following differential equation
\begin{align}
\frac{d}{ds}b_{nl}^{(0)}(s)+\tilde{\Gamma}_{ll}(s)b_{nl}^{(0)}(s) & =0\label{eq:zero order}\\
\frac{d}{ds}b_{nl}^{(1)}(s)+\tilde{\Gamma}_{ll}(s)b_{nl}^{(1)}(s)+\sum_{m\ne l}\frac{d}{ds}\left(i\tilde{T}_{ml}\left(s\right)e^{-i\tau[\tilde{\phi}_{m}\left(s\right)-\tilde{\phi}_{l}(s)]}b_{nm}^{(0)}(s)\right) & =0.\label{eq:first order}
\end{align}
Here, $\tilde{T}_{ml}\left(s\right)=\tilde{\Gamma}_{lm}(s)/[\tilde{E}_{m}(s)-\tilde{E}_{l}(s)]$
denotes the non-adiabatic transition rate between the state $\left|\tilde{l}(s)\right\rangle $
and $\left|\tilde{m}(s)\right\rangle $.

According to the initial condition $c_{nl}(0)=\delta_{ln}$, we attain
the initial condition $b_{nl}^{(0)}(0)=\delta_{ln}$ and $b_{nl}^{(1)}(0)=0$
for Eq. (\ref{eq:zero order}) and Eq. (\ref{eq:first order}) respectively.
The solutions to Eqs (\ref{eq:zero order}) and (\ref{eq:first order})
follow as
\begin{align}
b_{nl}^{(0)}(s) & =\begin{cases}
0 & n\ne l\\
e^{i\tau\tilde{\gamma}_{n}(s)} & n=l
\end{cases}\\
b_{nl}^{(1)}(s) & =\begin{cases}
-i\left[\tilde{T}_{nl}(s)e^{-i\tau[\tilde{\phi}_{n}(s)-\tilde{\phi}_{l}(s))]+i\tilde{\gamma}_{n}(s)}-\tilde{T}_{nl}(0)e^{i\tilde{\gamma}_{l}(s)}\right] & n\ne l\\
0 & n=l
\end{cases},
\end{align}
with the Berry phase $\tilde{\gamma}_{l}(s)=i\int_{0}^{s}\tilde{\Gamma}_{ll}(s^{\prime})ds^{\prime}$.
In the main content, Equation (3) is obtained via $c_{nl}^{[1]}(\tau)=b_{nl}^{[1]}(1)$.
We remark that the current derivation of the adiabatic approximation
is the straightforward version. A more careful derivation can be found
in Ref. \citep{Rigolin_2008}, where the first-order result for $c_{nn}^{[1]}(\tau)$
contains a phase correction. Yet such phase has no effect on the absolute
square $\left|c_{nn}^{[1]}(\tau)\right|^{2}$ and in turn would not
change the results obtained from the current derivation.

For the initial thermal state, the work $W(\tau)=\sum_{n}p_{n}\left[\left\langle \psi_{n}(\tau)\right|H(\tau)\left|\psi_{n}(\tau)\right\rangle -\tilde{E}_{n}(0)\right]$
is given explicitly as 
\begin{equation}
W(\tau)=\sum_{n}p_{n}\left[\tilde{E}_{n}(1)-\tilde{E}_{n}(0)+\sum_{l\ne n}\left(\tilde{E}_{l}(1)-\tilde{E}_{n}(1)\right)\left|c_{nl}(\tau)\right|^{2}\right],\label{eq:worktau}
\end{equation}
Here, $p_{n}=\exp\left[-\beta\tilde{E}_{n}(0)\right]/\sum_{m}\exp\left[-\beta\tilde{E}_{m}(0)\right]$
denotes the initial thermal distribution with the inverse temperature
$\beta=1/(k_{B}T)$. For an quasi-static adiabatic process with long
control time $\tau\rightarrow\infty$, the solution by Eq. (3) in
the main content implies $\left|c_{nl}(\tau)\right|^{2}\rightarrow0,\,n\ne l$,
and the corresponding work approaches $W_{\mathrm{adi}}=\sum_{n}p_{n}\left[\tilde{E}_{n}(1)-\tilde{E}_{n}(0)\right]$.
The rest part of the work in Eq. (\ref{eq:worktau}) is named as the
extra work for the finite-time adiabatic process
\begin{equation}
W^{\mathrm{(ex)}}(\tau)=\sum_{n}p_{n}\left[\sum_{l\ne n}\left(\tilde{E}_{l}(1)-\tilde{E}_{n}(1)\right)\left|c_{nl}(\tau)\right|^{2}\right],\label{eq:ex work}
\end{equation}
which is Eq. (\ref{eq:extraworkwithc}) in the main content.

\section{1D Quantum Piston Model\label{sec:1DPistonModel}}

In this appendix, we show the details about the realization of the
finite-time quantum Otto cycle with 1D quantum piston model. Explicit
results of the maximal power and the EMP are derived for this model.

\subsection{$\mathcal{C}/\tau^{2}$ scaling of the extra work}

First, we show the $\mathcal{C}/\tau^{2}$ scaling of the extra work
for 1D quantum piston model during the finite-time  adiabatic process.
The time-dependent Hamiltonian $H(t)$ is given by Eq. (\ref{eq:timedependenthamiltonian})
of the main content with the control protocol $\tilde{L}(s)=L_{0}+(L_{1}-L_{0})s$.The
instantaneous wave-function and the corresponding energy $\tilde{E}_{n}(s)$
are given by Eqs. (\ref{eq:instantaneouswavefunction}) and (\ref{eq:correspondingenergy}).
$\tilde{\Gamma}_{ln}(s)$ of this model follows explicitly as 
\begin{align}
\tilde{\Gamma}_{ll}(s) & =0,\\
\tilde{\Gamma}_{ln}(s) & =\frac{2nl(-1)^{l+n}\left(L_{1}-L_{0}\right)}{\left(l^{2}-n^{2}\right)\tilde{L}(s)},\,l\ne n.
\end{align}
Therefore, the Berry phase vanishes in this model, namely $\tilde{\gamma_{l}}=0$.
And the non-adiabatic transition rate is 
\begin{equation}
\tilde{T}_{nl}(s)=-\frac{4Mnl(-1)^{l+n}\left(L_{1}-L_{0}\right)}{\pi^{2}(n^{2}-l^{2})^{2}}\tilde{L}(s).
\end{equation}
Substituting the rate into Eq. (3) in the main content, we obtain
the amplitude explicitly as

\begin{equation}
c_{nl}^{[1]}(\tau)=i\frac{4Mnl(-1)^{l+n}\left(L_{1}-L_{0}\right)}{\tau\pi^{2}(n^{2}-l^{2})^{2}}\left(L_{1}e^{-i\tau\frac{\pi^{2}\left(n^{2}-l^{2}\right)}{2ML_{0}L_{1}}}-L_{0}\right).
\end{equation}

By summing over the initial thermal distribution, we obtain the explicit
result for the extra work

\begin{equation}
W^{\mathrm{(ex)}}(\tau)=W^{\mathrm{(mean)}}(\tau)+W^{\mathrm{(osc)}}(\tau),\label{eq:exwork thermal piston}
\end{equation}
where the mean extra work is

\begin{equation}
W^{\mathrm{(mean)}}(\tau)=\frac{ML_{1}^{2}}{\tau^{2}}\left(1-r\right)^{2}\left(1+r^{2}\right)\left(\frac{1}{6}-\sum_{n=1}^{\infty}\frac{p_{n}}{4\pi^{2}n^{2}}\right),\label{eq:nmean}
\end{equation}
and the oscillating extra work is

\begin{equation}
W^{\mathrm{(osc)}}(\tau)=-\sum_{n=1}^{\infty}p_{n}\frac{16ML_{1}^{2}\left(1-r\right)^{2}r}{\tau^{2}}\sum_{l\ne n}\frac{l^{2}n^{2}}{\pi^{2}(l^{2}-n^{2})^{3}}\cos\left(\frac{\tau\left(n^{2}-l^{2}\right)\pi^{2}}{2MrL_{1}^{2}}\right),\label{eq:nosc}
\end{equation}
where $r=L_{0}/L_{1}$ is the expansion ratio, and $p_{n}=p_{n}(\beta,L_{0})$
is the initial thermal distribution given by Eq. (\ref{eq:pistoninitialthermal distribution})
in the main content. With Eqs. (\ref{eq:nmean}) and (\ref{eq:nosc}),
the coefficients in Eq. (\ref{eq:meancoe}) and Eq. (\ref{eq:osccoe})
of the main content is written explicitly as 
\begin{equation}
\Sigma=ML_{1}^{2}\left(1-r\right)^{2}\left(1+r^{2}\right)\left(\frac{1}{6}-\sum_{n=1}^{\infty}\frac{p_{n}}{4\pi^{2}n^{2}}\right),\label{eq:sigma}
\end{equation}
and 
\begin{equation}
\omega(\tau)=-16ML_{1}^{2}\left(1-r\right)^{2}r\sum_{n=1}^{\infty}\sum_{l\ne n}p_{n}\frac{l^{2}n^{2}}{\pi^{2}(l^{2}-n^{2})^{3}}\cos\left(\frac{\tau\left(n^{2}-l^{2}\right)\pi^{2}}{2MrL_{1}^{2}}\right).\label{eq:omegatau}
\end{equation}
For high temperature with the thermal de Broglie wavelength $\lambda_{\mathrm{th}}=\sqrt{2\pi\beta/M}$
much smaller than the length $L_{0}$ of the box, the summation by
Eq. (\ref{eq:sigma}) can be approximated as $p_{n}/(4\pi^{2}n^{2})\approx\int_{n-1/2}^{n+1/2}p_{n}/(4\pi^{2}n^{2})dn$.
And the summation over the index $n$ can be estimated as 
\begin{align}
\sum_{n=1}^{\infty}\frac{p_{n}(\beta,L_{0})}{4\pi^{2}n^{2}} & \approx\int_{1/2}^{\infty}\frac{1}{4\pi^{2}n^{2}}\frac{\exp\left(-\frac{\beta\pi^{2}n^{2}}{2ML_{0}^{2}}\right)}{\sqrt{\frac{L_{0}^{2}M}{2\pi\beta}}\text{erfc}\left(\sqrt{\frac{\beta\pi^{2}}{8L_{0}^{2}M}}\right)}dn\\
 & =\frac{e^{-\frac{\pi^{2}\beta}{8L_{0}^{2}M}}\sqrt{\frac{\beta}{2\pi^{3}L_{0}^{2}M}}}{\text{erfc}\left(\sqrt{\frac{\beta\pi^{2}}{8L_{0}^{2}M}}\right)}-\frac{\beta}{4L_{0}^{2}M}\\
 & =\sqrt{\frac{\beta}{2\pi^{3}L_{0}^{2}M}}+O(\beta).
\end{align}
Therefore, we neglect the last summation term in Eq. (\ref{eq:sigma})
at high temperature limit and simplify both the coefficient $\Sigma$
as 
\begin{equation}
\Sigma=\frac{M}{6}ML_{1}^{2}\left(1-r\right)^{2}\left(1+r^{2}\right),\label{eq:SigmahighT}
\end{equation}
and the approximate mean extra work $W^{(\mathrm{mean})}$ is given
by Eq. (\ref{eq:wmeanpistonappro}) in the main content.

\subsection{The Exact Solution}

The current model can be solved analytically as shown in Ref. \citep{Doescher_1969,Makowski_1991,Quan2012}.
Here, we only show the relevant part of the exact solution for the
later numerical calculations. For the given protocol above, the exact
solution for the time-dependent Schroedinger equation $i\partial_{t}\left|\Psi_{n}(t)\right\rangle =H(t)\left|\Psi_{n}(t)\right\rangle $
exists
\begin{equation}
\left\langle x\left|\Psi_{n}(t)\right.\right\rangle =e^{i\left(\frac{1}{2}Mx^{2}\frac{L_{1}-L_{0}}{L(t)\tau}-\frac{n^{2}\pi^{2}}{2ML_{0}L(t)}t\right)}\sqrt{\frac{2}{L(t)}}\sin\frac{n\pi x}{L(t)},
\end{equation}
with $L(t)=L_{0}+\left(L_{1}-L_{0}\right)t/\tau$. Here, the time-dependent
solution $\left|\Psi_{n}(t)\right\rangle $ forms a complete orthogonal
set at any given time $t$. Therefore, the initial eigen-state $\left|\psi_{n}(0)\right\rangle =\left|n(0)\right\rangle $
can be expanded with $\left|\Psi_{l}(0)\right\rangle $ as 
\begin{equation}
\left|n(0)\right\rangle =\sum_{l=1}^{\infty}\left\langle \Psi_{l}(0)\left|n(0)\right.\right\rangle \left|\Psi_{l}(0)\right\rangle ,
\end{equation}
and the state at time $\tau$ follows as 
\begin{equation}
\left|\psi_{n}(\tau)\right\rangle =\sum_{l=1}^{\infty}\left\langle \Psi_{l}(0)\left|n(0)\right.\right\rangle \left|\Psi_{l}(\tau)\right\rangle .
\end{equation}
 For an initial thermal state with the distribution $p_{n}=p_{n}(\beta,L_{0})$,
the work is determined by the change of the internal energy
\begin{equation}
W(\tau)=\sum_{n=1}^{\infty}p_{n}\left(\sum_{l=1}^{\infty}\left|\left\langle \Psi_{l}(0)\left|n(0)\right.\right\rangle \right|^{2}\left\langle \Psi_{l}(\tau)\right|H(\tau)\left|\Psi_{l}(\tau)\right\rangle -E_{n}(0)\right)
\end{equation}
with $H(t)$ given by Eq. (\ref{eq:timedependenthamiltonian}) in
the text and the initial energy $E_{n}(0)=n^{2}\pi^{2}/\left(2ML_{0}^{2}\right)$.
The extra work follows from Eq. (\ref{eq:ex work}) as 
\begin{equation}
W^{(\mathrm{ex})}(\tau)=\sum_{n=1}^{\infty}p_{n}\left(\sum_{l=1}^{\infty}\left|\left\langle \Psi_{l}(0)\left|n(0)\right.\right\rangle \right|^{2}\left\langle \Psi_{l}(\tau)\right|H(\tau)\left|\Psi_{l}(\tau)\right\rangle -E_{n}(\tau)\right).
\end{equation}
The exact extra work is obtained by numerically calculating the initial
projection $\left\langle \Psi_{l}(0)\left|n(0)\right.\right\rangle $
and the internal energy $\left\langle \Psi_{l}(\tau)\right|H(\tau)\left|\Psi_{l}(\tau)\right\rangle $.

\subsection{The Validation of the Scaling}

With the exact solution, we can validate the obtained $\mathcal{C}/\tau^{2}$
scaling. In addition to the expansion process (Fig. 3(a) in the main
content), we supplement the $\mathcal{C}/\tau^{2}$ scaling of the
extra work in the compression process in Fig. \ref{fig1}(a). We set
the mass and the Boltzmann constant as $M=1$, $k_{B}=1$, and consider
three initial thermal equilibrium states with the temperature $T=1,\,50,\,100$
(blue circle, black square, and red diamond in Fig. \ref{fig1} (a)).
For higher temperature, the oscillation of the extra work becomes
weaker. And the exact numerical results matches the mean extra work
in Eq. (\ref{eq:wmeanpistonappro}) at high temperature (shown as
the green line). In Fig. \ref{fig1}(b), we compare the total extra
work (the curves) in Eq. (\ref{eq:exwork thermal piston}), with the
exact numerical results (the markers). The curves show a good match
with the exact numerical results for both the compression and the
expansion processes with long control time $\tau$.

\begin{figure}
\includegraphics[width=8cm]{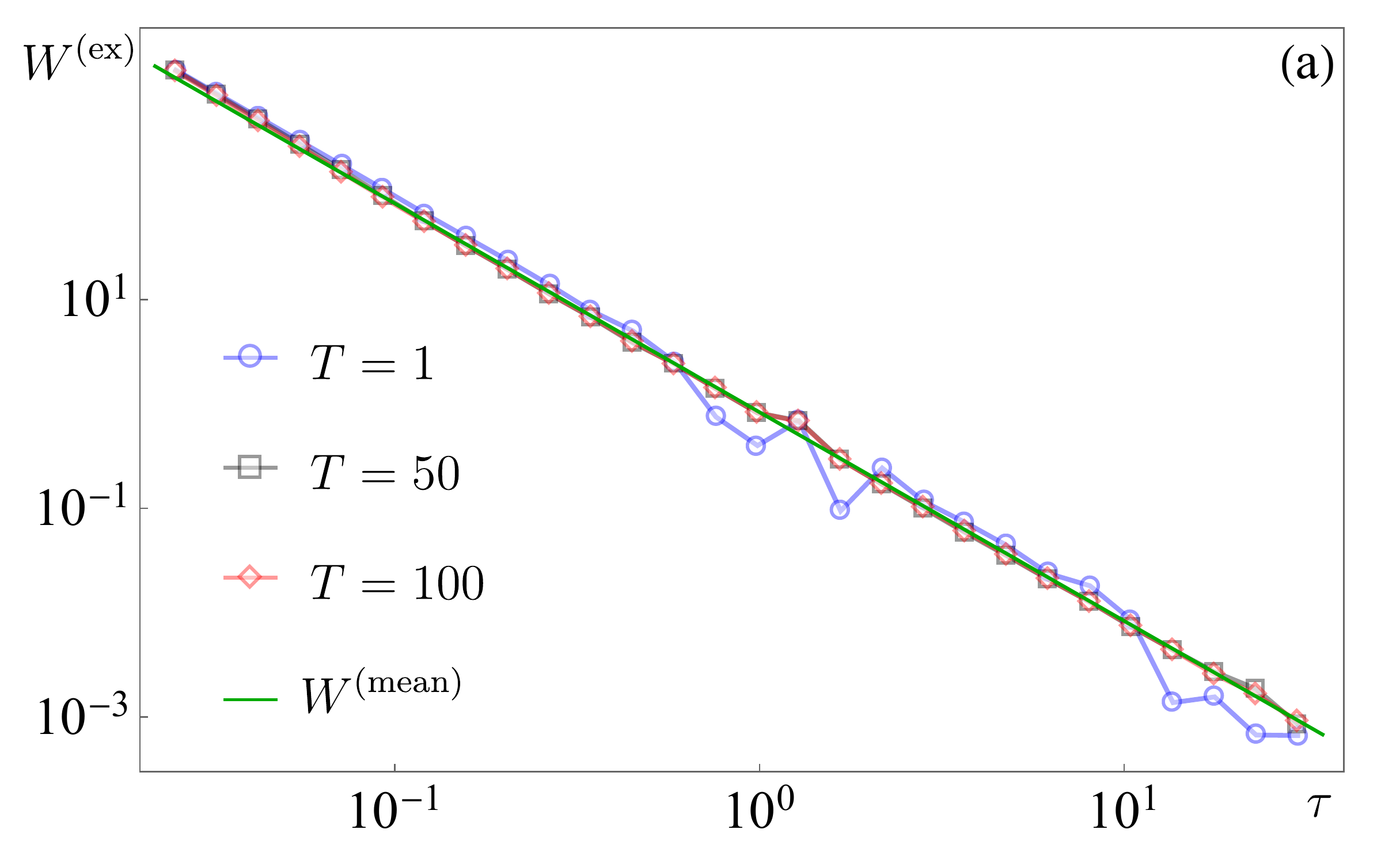}\includegraphics[width=8cm]{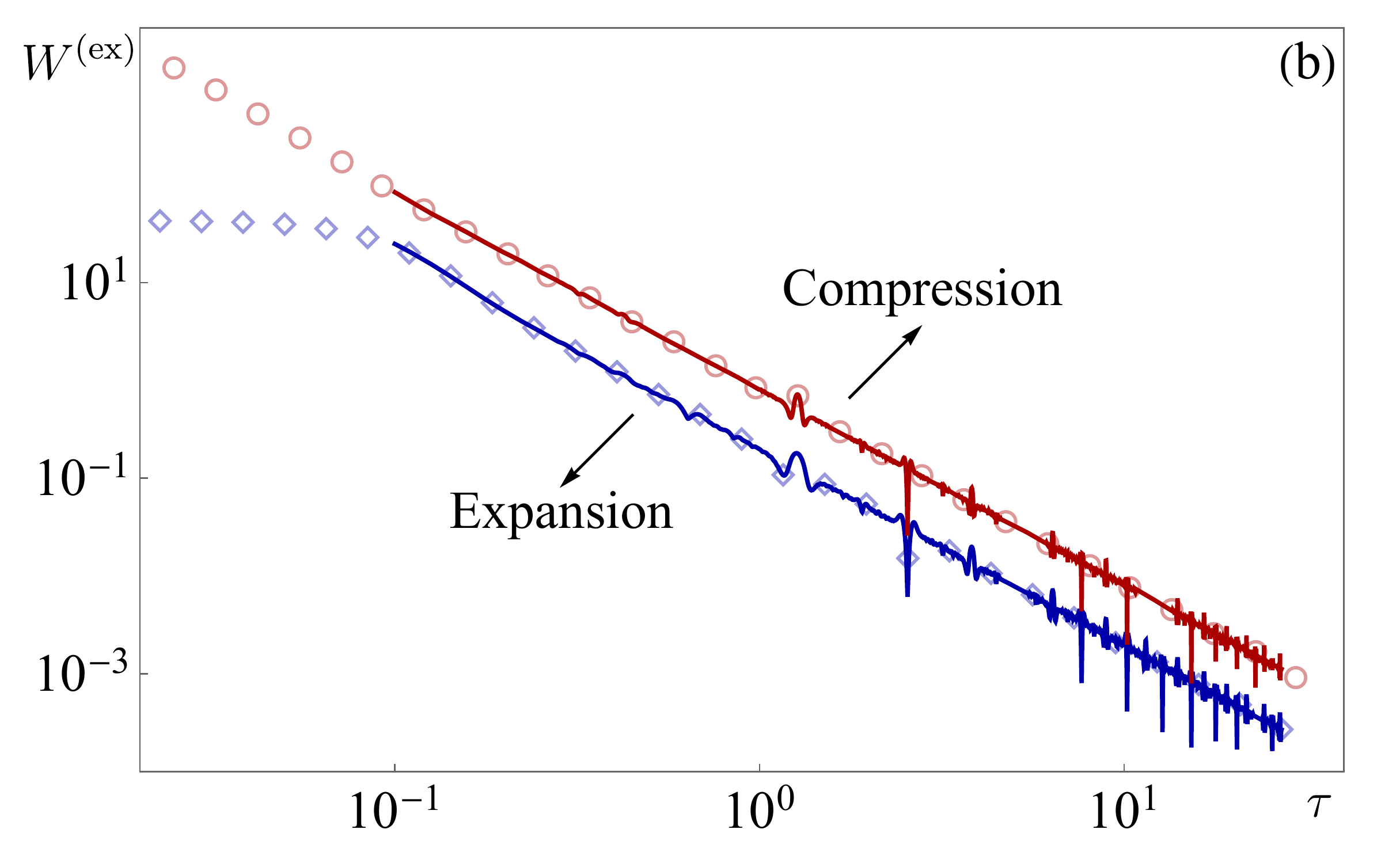}\caption{(a) The $\mathcal{C}/\tau^{2}$ scaling for the extra work in the
finite-time adiabatic compression process. The length of the box decreases
from $L_{0}=2$ to $L_{1}=1$. The exact numerical results for three
initial thermal equilibrium states with temperatures $T=1,\,50,\,100$
are presented in the blue circles, black squares, and red diamonds.
And the green line shows the mean extra work $W^{(\mathrm{mean})}$
in Eq. (\ref{eq:wmeanpistonappro}). (b) The extra work for the expansion
and compression process. The length of the box changes from $L_{0}=2(1)$
to $L_{1}=1(2)$ for the compression (expansion) process at the temperature
$T=100$. The upper red (lower blue) line with markers present the
total extra work for compression (expansion) process. The markers
show the exact numerical results while the line is obtained by Eq.
(\ref{eq:exwork thermal piston}).}

\label{fig1}
\end{figure}

\subsection{The Engine Cycle}

To optimize a finite-time Otto heat engine, we need the net work $W_{\mathrm{T}}^{\mathrm{adi}}$
and the efficiency $\eta^{\mathrm{adi}}$ for the quasi-static Otto
cycle. Considering the Otto cycle given in Fig. 1(b) in the main content,
the internal energy for the equilibrium state $1$ and $3$ is $\mathrm{Tr}\left[\rho_{1}H_{1}\right]=\sum_{n=1}^{\infty}p_{n}^{(1)}\tilde{E}_{n}(0)$
and $\mathrm{Tr}\left[\rho_{3}H_{3}\right]=\sum_{n=1}^{\infty}p_{n}^{(3)}\tilde{E}_{n}(1)$
with the thermal distribution $p_{n}^{(1)}=p_{n}(\beta_{h},L_{0})$
and $p_{n}^{(3)}=p_{n}(\beta_{c},L_{1})$ . For a quasi-static Otto
cycle, the distribution during the quasi-static adiabatic processes
remains its initial distribution, which leads to the internal energy
for the state $2$ and $4$ as $\mathrm{Tr}\left[\rho_{2}H_{2}\right]=\sum_{n=1}^{\infty}p_{n}^{(1)}\tilde{E}_{n}(1)$
and $\mathrm{Tr}\left[\rho_{4}H_{4}\right]=\sum_{n=1}^{\infty}p_{n}^{(3)}\tilde{E}_{n}(0)$.
We obtain the heat absorbed from the hot bath as 
\begin{equation}
Q_{h}^{\mathrm{adi}}=\sum_{n=1}^{\infty}(p_{n}^{(1)}-p_{n}^{(3)})\tilde{E}_{n}(0),
\end{equation}
and the net work as 
\begin{equation}
W_{\mathrm{T}}^{\mathrm{adi}}=\sum_{n=1}^{\infty}(p_{n}^{(1)}-p_{n}^{(3)})\left(\tilde{E}_{n}(1)-\tilde{E}_{n}(0)\right).\label{eq:Wadipiston}
\end{equation}
The efficiency for the quasi-static Otto cycle follows as 
\begin{equation}
\eta^{\mathrm{adi}}=1-r^{2}\label{eq:etaadipiston}
\end{equation}
with the ratio $r=L_{0}/L_{1}$. At high temperature, the summation
in Eq. (\ref{eq:Wadipiston}) can be approximated as 
\begin{equation}
W_{\mathrm{T}}^{\mathrm{adi}}\approx\frac{k_{B}}{2}\left(T_{h}r^{2}-T_{c}\right)\frac{1-r^{2}}{r^{2}}.\label{eq:Wadipistonhigh}
\end{equation}

For the current model with 1D quantum piston, the extra work at high
temperature is

\begin{align}
\Sigma_{1} & =\frac{ML_{1}^{2}}{6}\left(1-r\right)^{2}\left(1+r^{2}\right),\label{eq:sigma1piston}
\end{align}
and 
\begin{equation}
\Sigma_{3}=\frac{ML_{1}^{2}}{6r^{2}}\left(1-r\right)^{2}\left(1+r^{2}\right).\label{eq:sigma3piston}
\end{equation}
With the explicit result of Eqs (\ref{eq:etaadipiston})-(\ref{eq:sigma3piston}),
the optimal control time follows as 
\begin{equation}
\tau_{1}^{*}=\sqrt{\frac{ML_{1}^{2}\left(1-r\right)\left(1+r^{2}\right)(r^{4/3}+r^{2})}{k_{B}\left(T_{h}r^{2}-T_{c}\right)\left(1+r\right)}},\label{eq:tau1op-1}
\end{equation}
and
\begin{equation}
\tau_{3}^{*}=\sqrt{\frac{ML_{1}^{2}\left(1-r\right)\left(1+r^{2}\right)\left(r^{2/3}+1\right)}{k_{B}\left(T_{h}r^{2}-T_{c}\right)\left(1+r\right)}}.
\end{equation}
And the maximal power and the EMP by Eqs. (\ref{eq:maxpower}) and
(\ref{Emp}) is obtained as 
\begin{equation}
P_{\mathrm{max}}^{\mathrm{Piston}}=\frac{1}{3L_{1}}\left[\frac{k_{B}\left(T_{h}r^{2}-T_{c}\right)\left(1-r^{2}\right)}{\left(M\left(1-r\right)^{2}\left(1+r^{2}\right)\right)^{1/3}\left(r^{2}+r^{4/3}\right)}\right]^{\frac{3}{2}},
\end{equation}
and 
\begin{equation}
\eta_{\mathrm{EMP}}^{\mathrm{Piston}}=\frac{2\left(1-r^{2}\right)}{3-(1-r^{2})/(1+r^{2/3})}.
\end{equation}
The efficiency can be rewritten with the quasi-static efficiency $\eta^{\mathrm{adi}}$
as 
\begin{equation}
\eta_{\mathrm{EMP}}^{\mathrm{Piston}}=\frac{2\eta^{\mathrm{adi}}}{3-\eta^{\mathrm{adi}}/[(1-\eta^{\mathrm{adi}})^{1/3}+1]},
\end{equation}
which is Eq. (\ref{eq:emppiston}) in the main content.

\begin{figure}
\includegraphics[width=5cm]{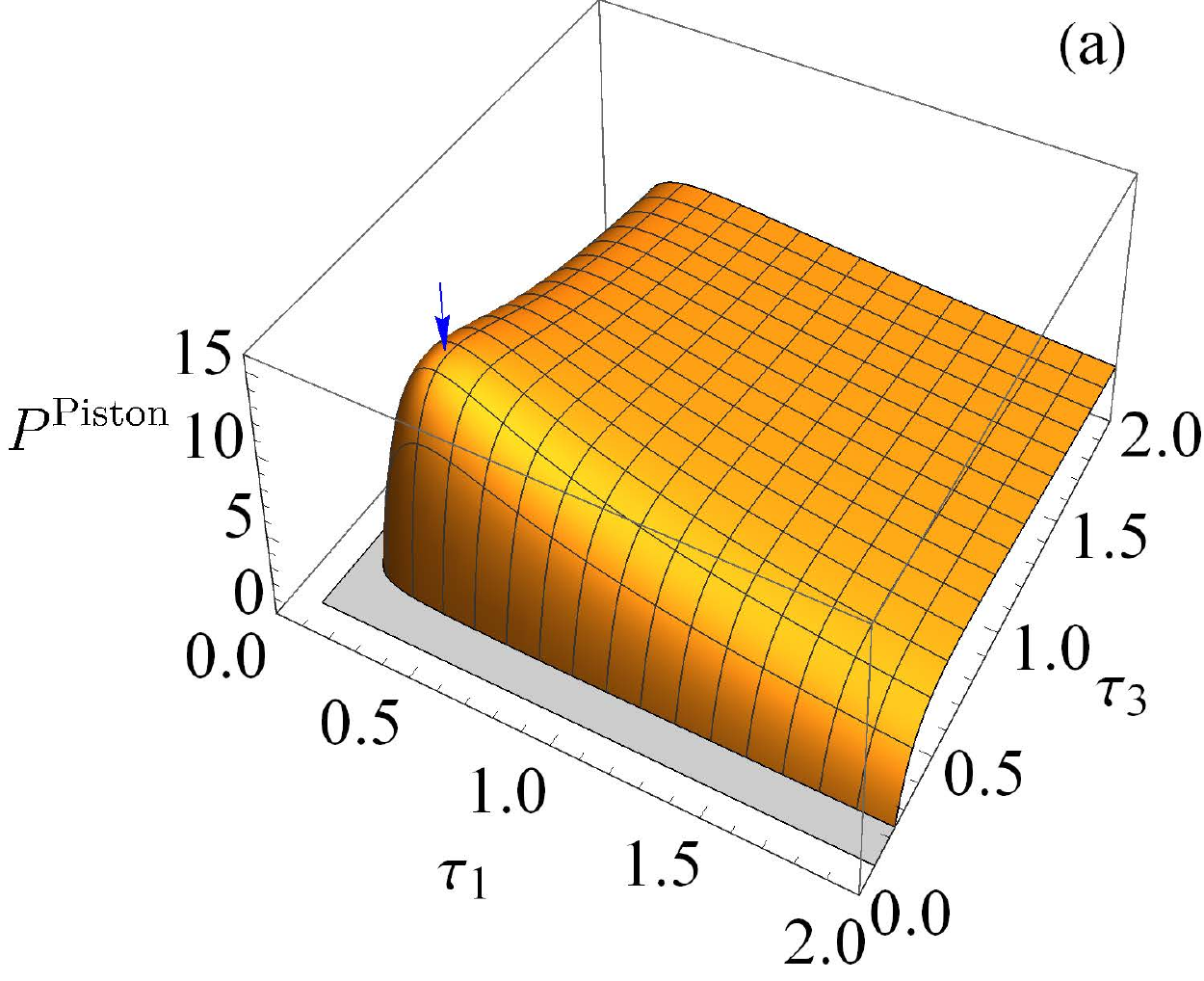}\,\,\includegraphics[width=5cm]{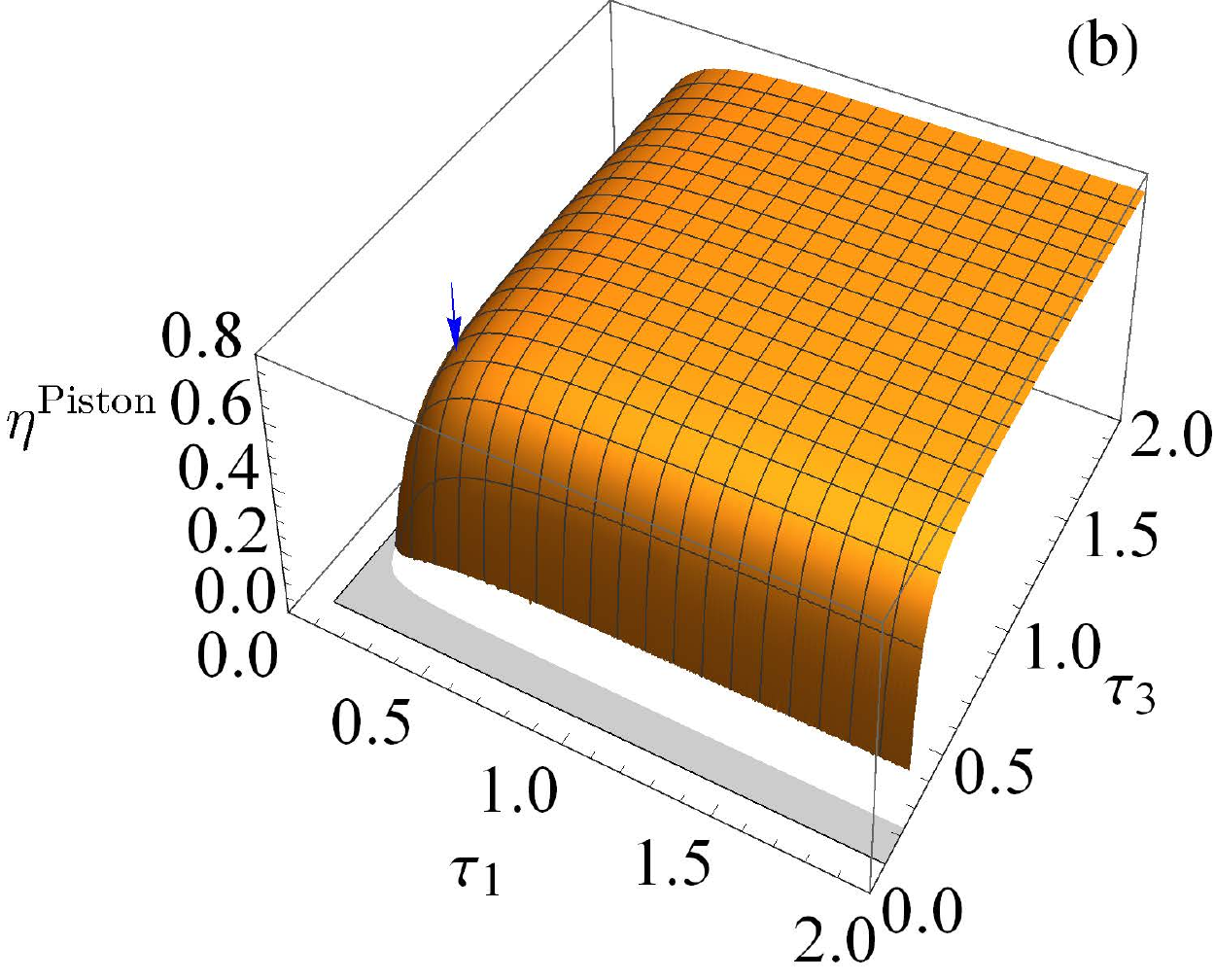}\,\,\,\,\,\,\,\,\includegraphics[width=6cm]{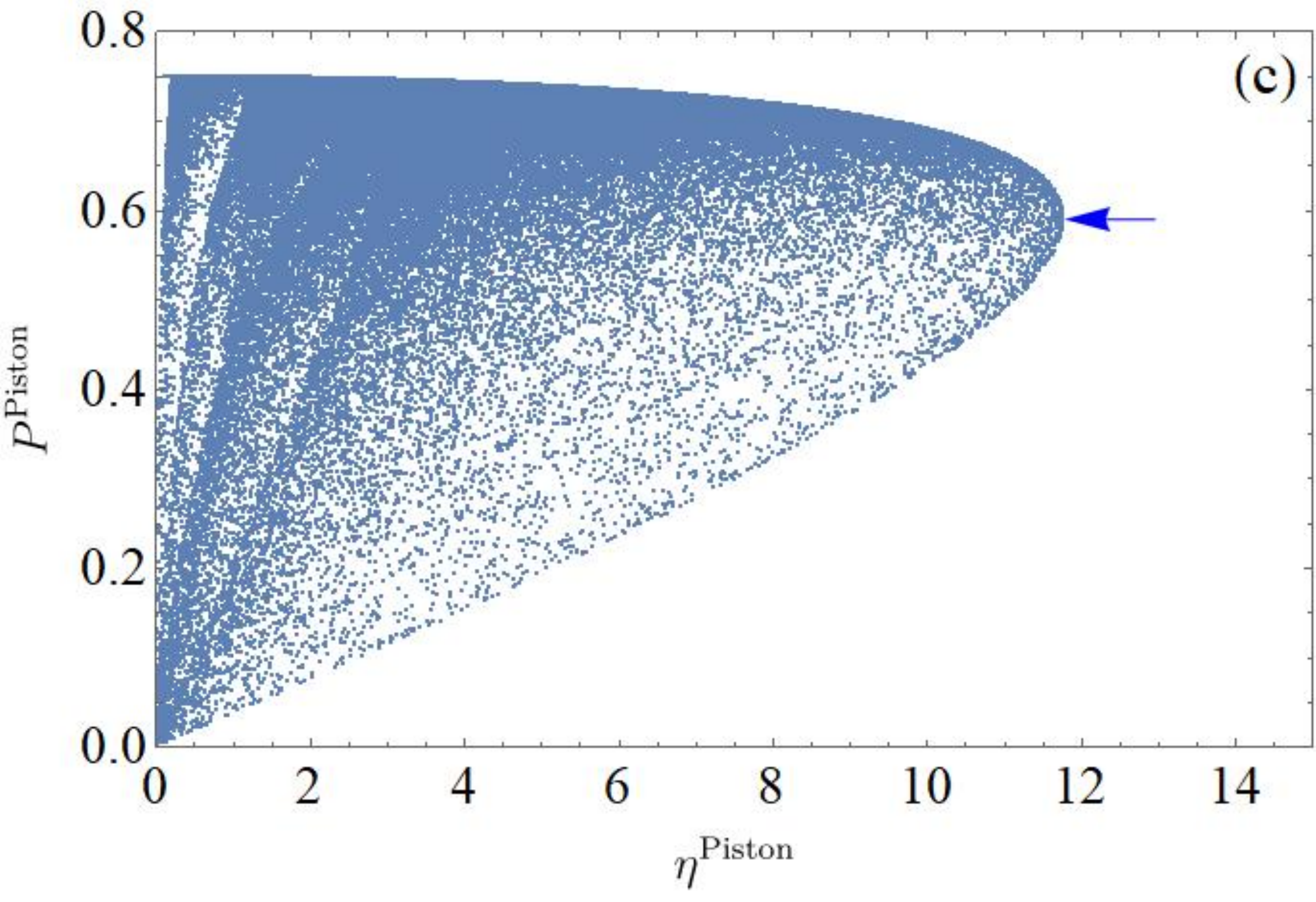}

\caption{(a) The power and (b) the efficiency as function of the two control
time $\tau_{1}$ and $\tau_{3}$ in the 1D quantum piston model. The
parameters are chosen as $T_{h}=100,T_{c}=20,L_{0}=1$ and $L_{1}=2$.
The power has a maximum for particular $\tau_{1}^{*}$ and $\tau_{3}^{*}$.
The blue arrows gives the maximum power in subfigure (a) and the EMP
in subfigure (b). (c) The achievable $(P,\eta)$ for different $\tau_{1}$
and $\tau_{3}$. The blue arrow gives the maximum power and the corresponding
EMP.}

\label{fig2-1}
\end{figure}

In Fig. \ref{fig2-1}(a) and (b), we show the efficiency and the power
for the finite-time quantum Otto cycle as functions of the control
time $\tau_{1},\tau_{3}$ of the finite-time  adiabatic process. Fig.
\ref{fig2-1}(a) shows that the power has a maximum power at the particular
control time $\tau_{1}^{*},\tau_{3}^{*}$ (marked with the blue arrow).
The corresponding EMP in Fig. \ref{fig2-1}(b) is given with the blue
arrow. In Fig. \ref{fig2-1}(c), we show the constraint between the
power and the efficiency, by randomly choosing 600,000 pairs of $(\tau_{1},\tau_{3})$
to calculate the corresponding power and efficiency. A clear bound
appears, which shows a cutoff between the power and and the efficiency.
The maximum power along with the EMP is marked with the blue arrow.
The detailed discussion of the exact constraint relation will be presented
elsewhere.

\end{widetext}
\end{document}